\documentclass[aps,prl,twocolumn,superscriptaddress]{revtex4}
\usepackage[dvips]{graphicx}
\usepackage{amsmath}

\renewcommand{\bar}[1]{\overline{#1}}
\newcommand{\half}{{\frac{1}{2}}}
\newcommand{\threehalf}{{\frac{3}{2}}}
\newcommand{\fivehalf}{{\frac{5}{2}}}

\newcommand{\ninehalf}{{\frac{9}{2}}}

\def\Dslash{\raise.15ex\hbox{/}\kern-.7em D}
\def\Pslash{\raise.15ex\hbox{/}\kern-.7em P}

\begin{document}

\preprint{NSF-KITP-04-131}
\preprint{SLAC-PUB-10789}

\title{Hadronic Spectrum of a Holographic Dual of QCD}

\author{Guy F. de T\'eramond}
\affiliation{Universidad de Costa Rica, San Jos\'e, Costa Rica}
\author{Stanley J. Brodsky}
\affiliation{Stanford Linear Accelerator Center, Stanford University,
Stanford, California 94309, USA}


\begin{abstract}

We compute the spectrum of light hadrons in a holographic dual of QCD
defined on $AdS_5 \times S^5$
which has conformal
behavior at short distances and confinement at large
interquark separation.
Specific hadrons are identified by the correspondence of
string modes with the
dimension of the interpolating operator of the hadron's valence Fock
state. Higher orbital excitations are matched quanta to quanta
with fluctuations about the AdS background.
Since only one parameter, the
QCD scale $\Lambda_{\rm QCD}$, is used, the agreement with
the pattern of physical states is remarkable. In particular,
the ratio of delta to nucleon trajectories is determined by
the ratio of zeros of Bessel functions.

\end{abstract}


\maketitle

The correspondence~\cite{Maldacena:1997re} between
10-dimensional string theory defined on anti--de Sitter ($AdS_5 \times S^5$)
and Yang-Mills theories at its conformal 3+1
space-time boundary~\cite{Gubser:1998bc} has led to important insights
into the properties of QCD at strong coupling.
As shown by Polchinski
and Strassler~\cite{Polchinski:2001tt},  one can
give a nonperturbative derivation of  dimensional counting
rules~\cite{Brodsky:1973kr} for the leading
power-law fall-off of hard exclusive glueball scattering in
gauge theories with a mass gap dual to supergravity in warped
space-times.  The resulting theories
have the hard behavior expected from QCD at short distances, rather than the soft
behavior characteristic of string theory.  Another important application
is the description of
deep inelastic scattering structure functions at small
$x$~\cite{Polchinski:2002jw}.
One can also derive the falloff of
hadronic light-front wavefunctions in QCD at large transverse momentum by matching
their short-distance properties to the
behavior of the string solutions in the large-$r$ conformal region of AdS
space~\cite{Brodsky:2003px}. The scale dependence of the string modes
determines the analytic behavior
of the QCD hadronic wave function, providing a precise counting rule
for each Fock component with
any number of quarks and gluons and any internal orbital
angular momentum~\cite{Brodsky:2003px}.
The predicted orbital dependence coincides with perturbative
QCD results~\cite{Ji:2003fw}.

The $\mathcal{N} = 4$ super Yang-Mills (SYM) theory
at large $N_C$ in four dimensions is dual to the low energy
supergravity approximation to type IIB string
compactified on $AdS_5 \times S^5$~\cite{Maldacena:1997re}.
However, QCD is fundamentally different from SYM
theories where all of the matter fields appear in adjoint multiplets of
$SU(N_C)$. The introduction of quarks in the fundamental representation is dual to
the introduction of an open string
sector~\cite{Gross:1998gk}. 

There is now substantial
theoretical~\cite{Alkofer:2004it}  and empirical~\cite{Brodsky:2004qb}  evidence
that the QCD coupling has an IR
fixed point.  In many phenomenological  applications, such as exclusive
processes at experimentally accessible momentum transfers, the amplitudes
are evaluated in the regime where the exchanged gluon momenta are not
very large and the QCD coupling is nearly constant and not small
~\cite{Brodsky:1997dh}.  
For example, the phenomenological successes of dimensional counting rules for
exclusive processes can be understood if QCD
resembles a strongly coupled conformal theory at moderate but not
asymptotic momentum transfer. QCD is also a
confining gauge theory in the infrared with a mass gap $\Lambda_{\rm QCD}$
and a well-defined spectrum of color-singlet hadronic
states. 

The isomorphism  of the  group
$SO(2,4)$ of conformal QCD in the limit of
massless quarks and vanishing $\beta$ function
with the isometries of AdS space, $x^\mu \to \lambda x^\mu$, $r \to r/\lambda$,
maps scale transformations into the holographic coordinate $r$:
the string mode  in  $r$ is the extension
of the  hadron wave function into the fifth dimension.
Different values of $r$ correspond to
different energy scales at which the hadron is examined, and
determines the scale of the invariant
separation between quarks $x_\mu x^\mu \to \lambda^2 x_\mu x^\mu$.
In particular, the $r \to \infty$ boundary corresponds to the $Q \to
\infty$, zero separation limit.
Conversely, color confinement implies that there is a maximum separation of quarks
and a minimum value
of $r$. Thus, AdS space should end at a finite value
$r_0 = \Lambda_{\rm QCD} R^2$ truncating the regime where the string modes can propagate.
The cutoff at $r_0$ breaks conformal invariance
and allows the introduction of the QCD scale.

A 10-dimensional nonconformal metric
dual to a confining gauge theory is written
as~\cite{Polchinski:2001tt}
\begin{equation}
ds^2 = \frac{R^2}{z^2} e^{2A(z)} \left(\eta_{\mu \nu} dx^\mu dx^\nu - dz^2\right)
       + ds^2_X ,
\label{eq:zmetric}
\end{equation}
where $A(z) \to 0$ as $z = R^2/r \to 0$, and $R$ is the AdS radius.
The metric (\ref{eq:zmetric}) behaves asymptotically as
a product of AdS space and a compact manifold $X$.
Color confinement will be
described in a simplified model based on a ``hard-wall" approximation where
the metric factor $e^{2A(z)}$ is a step function.  This provides an
analog of the MIT bag model where quarks are permanently
confined inside a finite region of space~\cite{Chodos:1974je}. As in
the bag model the linearized equations in the bulk have no
interactions. However, unlike bag models,
the truncated boundary conditions on string modes are imposed on the
holographic coordinate, not on the bag wave function at fixed time.  The
truncated anti--de Sitter/conformal field theory (CFT) thus provides 
a manifestly Lorentz invariant model with
confinement at large distances and conformal behavior at short distances.

The AdS/CFT correspondence can be interpreted in the present context as a
classical duality between
the valence state of a hadron in the asymptotic $3 + 1$
boundary and the lightest mass string mode in
$AdS_5 \times S^5$~\cite{Brodsky:2003px,deTeramond:2004qd}.
Higher Fock components are manifestations of the
quantum fluctuations of QCD; metric
fluctuations of the bulk geometry about the fixed AdS background should
correspond to quantum fluctuations of Fock states above the valence state.
In fact, as shown by Gubser, Klebanov and Polyakov for large Lorentz spin,
orbital excitations in the boundary
correspond to string degrees of freedom propagating in the bulk
from quantum fluctuations in the AdS
sector~\cite{Gubser:2002tv}. We identify the higher spin hadrons with the
fluctuations around the spin 0, $\half$, 1 and $\threehalf$ string
solutions on $AdS_5$. This identification avoids the huge string
dimensions associated with spin $> 2$, which grow as $\Delta \sim
(g_s N_C)^\frac{1}{4}$ at large $N_C$.
The interpolating operators ${\cal O}$,
$\langle P \vert {\cal O} \vert 0 \rangle \neq 0$, which couple to the color-singlet hadrons
at the boundary can be
constructed from gauge-invariant products of local
quark and gluon fields taken at
the same point in four-dimensional space-time.
We introduce quarks
in the fundamental representation
at the AdS boundary, and follow their wavefunctions as they
propagate into the bulk. 

As a first application of our procedure,
consider the twist (dimension minus spin) two glueball interpolating operators
$\mathcal{O}_{4 + L} = F D_{\{\ell_1} \dots D_{\ell_m\}} F$, written
in terms of the symmetrized product of covariant derivatives $D$. The
operator $\mathcal{O}_{4+L}$ has total internal space-time orbital
momentum, $L = \sum_{i=1}^m \ell_i$ and conformal dimension $\Delta = 4 + L$.
We shall match the large $r$ asymptotic behavior of each string mode
in the bulk to the corresponding
conformal dimension of the boundary operators
of each hadronic state while maintaining conformal invariance~\cite{deTeramond:2004qd}.
In the conformal limit, an $L$ quantum, which is
identified with a quantum fluctuation about the AdS geometry,
corresponds to an effective five-dimensional mass $\mu$ in
the bulk side.  The allowed values of $\mu$ are uniquely determined by
requiring that asymptotically the dimensions become spaced by
integers, according to the spectral relation $(\mu R)^2 =
\Delta(\Delta - 4)$.  For large space-time angular
momentum $L$, we recover the string theory  results for the spectrum of
oscillatory exited states $\mu \simeq L/R$.
The physical string modes are plane waves along the Poincar\'e
coordinates with four-momentum $P_\mu$ and hadronic
invariant mass states given by $P_\mu P^\mu = \mathcal{M}^2$.
The four-dimensional mass spectrum $\mathcal{M}_{L}$  follows from
the boundary condition $\Phi(x,z_o) = 0$ on the solutions of the AdS
wave equation with effective mass $\mu$:
\vspace{-1.0pt}
\begin{equation}
\left[  z^2 ~\partial_z^2 - (d - 1) z ~ \partial_z + z^2~\mathcal{M}^2
- (\mu R)^2 \right] f(z) = 0,
\end{equation}
where $\Phi(x,z) = e^{- i P \cdot x} f(z)$. The normalizable modes are
\begin{equation}
\Phi_{\alpha,k}(x,z) = C_{\alpha,k}~e^{-i P \cdot x} z^2
J_\alpha\left(z \beta_{\alpha,k} \Lambda_{QCD}  \right),
\end{equation}
with $C_{\alpha,k}$ a normalization constant, $\alpha = 2 + L$ and $\Delta = 4 + L$ for $d = 4$.
For small $z$, $\Phi$ scales as $z^{-\Delta}$,
where the scaling dimension $\Delta$ of the string mode has
the same dimension of the interpolating operator which creates
a hadron. The four-dimensional mass spectrum is
then determined by
the zeros of Bessel functions $\beta_{\alpha,k}:$
\begin{equation}
\mathcal{M}_{\alpha, k} = \beta_{\alpha,k} \Lambda_{\rm QCD}.
\label{eq:Mzero}
\end{equation}
A similar expression for the glueball spectrum follows from
considering a AdS slice with boundary conditions at some finite value
of $z$~\cite{Boschi-Filho:2002ta}.

\begin{figure}[h]
\includegraphics[angle=0,width=8.5cm]{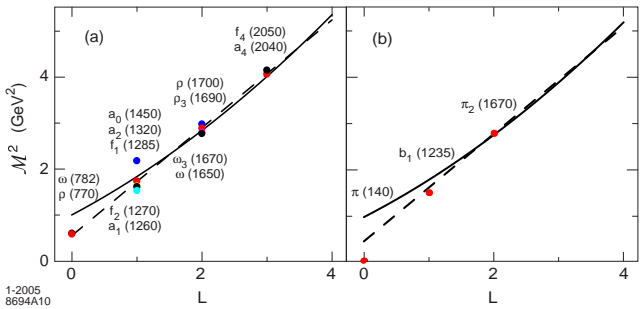}
\caption{Light meson orbital states for $\Lambda_{\rm QCD}$ = 0.263 GeV.
 Results for the vector mesons are shown in (a)
 and for the pseudoscalar mesons
 in (b). The dashed line corresponds to the  usual linear Regge
 trajectory and has slope 1.16 GeV$^2$.}
\label{fig:MesonSpec}
\end{figure}

We next consider the twist-two, dimension $3 + L$, vector-meson
operators
$\mathcal{O}_{3 + L}^\mu = \bar \psi \gamma^\mu D_{\{\ell_1} \dots
D_{\ell_m\}} \psi$, dual to string modes $\Phi_\mu = e^{- i P \cdot x}
f_\mu(z)$ propagating on AdS space with polarization along the Poincar\'e coordinates.
The string wavefunctions of the
vector mesons are then
determined by the five-dimensional wave equation
\begin{equation}
\left[z^2 \partial_z^2 - (d-1) z \partial_z + z^2 \mathcal{M}^2
- (\mu R)^2 +  d-1 \right] f_\mu(z) = 0,
\end{equation}
in the $\Phi_z = 0$ gauge~\cite{Muck:1998iz}, with normalizable modes
\begin{equation}
\Phi_{\alpha,k}^\mu(x,z) = C_{\alpha,k}
e^{-i P \cdot x} z^2
J_\alpha\left(z \beta_{\alpha,k} \Lambda_{QCD}  \right) \epsilon^\mu,
\end{equation}
where $\alpha = 1 + L$ and $\Delta = 3 + L$.  The hadronic mass
spectrum  follows from
$\Phi_\mu(x,z_o) = 0$.  Similarly, the
pseudoscalar mesons are described by the operator
$\mathcal{O}_{3 + L}= \bar \psi \gamma_5 D_{\{\ell_1} \dots
D_{\ell_m\}} \psi$, dual to string modes polarized along the radial
coordinate in the $\Phi_\mu = 0$ gauge.
The predicted spectrum is compared in Fig. \ref{fig:MesonSpec} with the
masses of light mesons listed by the Particle Data Group~\cite{Eidelman:2004wy}.
We plot the values of $\mathcal{M}^2$
as function of $L$ for
$\Lambda_{\rm QCD} = 0.263$ GeV. The predicted masses for the lightest hadrons
are too high, but otherwise the results are in good
agreement with the empirical values.
A string mode with a node in the
coordinate $r$ should correspond to a radial resonance with a
node in the interquark separation. The first radial AdS eigenvalue
has a mass 1.8 GeV which is high compared to the masses of the
observed radial excited mesons, the $\pi(1300)$ or the $\rho(1450)$.
These defects could possibly be cured by modifying the sharp cutoff at $r_0$.

The study of the baryon spectrum is crucial for our
understanding of bound states of strongly interacting relativistic
confined particles.
Consider the twist-three, dimension $\ninehalf + L$,
baryon operators
$\mathcal{O}_{(9/2) + L} =  \psi D_{\{\ell_1} \dots
D_{\ell_q } \psi D_{\ell_{q+1}} \dots D_{\ell_m\}} \psi$, dual to
spin-$\half$ or $\threehalf$ modes in the bulk. In this case,
we need to solve the full ten-dimensional Dirac wave equation,
$\Dslash \hat \Psi = 0$, since
the lowest Kaluza-Klein (KK) mode of the Dirac operator on an N-sphere
is not zero.  Consequently, baryons are charged under the $SU(4)_R
\sim SO(6)$ $R$ symmetry of $S^5$.  In contrast, the $SU(4)_R$ charge of mesons is zero.
We have classified the baryonic states according to the $SU(2)_F \otimes
SU(2)_{spin}  \subset  SU(4)$ isospin-spin symmetry corresponding to two
massless quarks. 

\begin{figure}[h]
\includegraphics[angle=0,width=7.375cm]{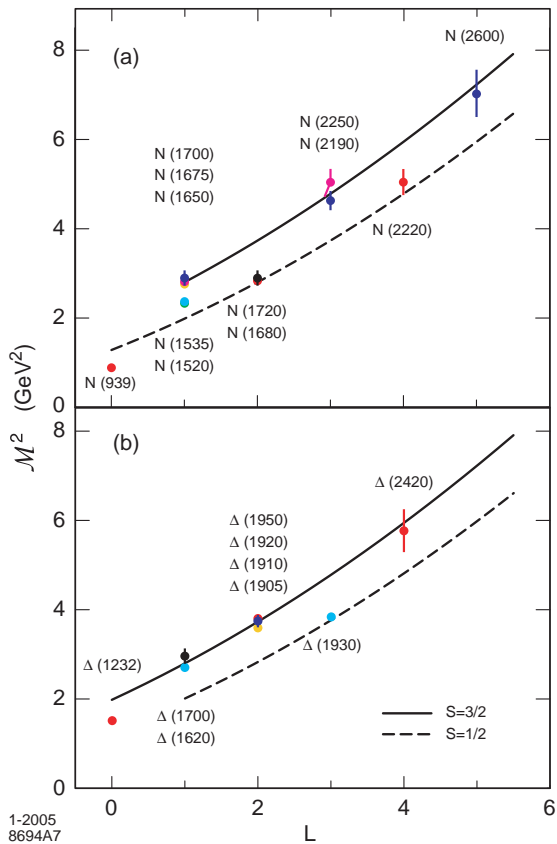}
\caption{Light baryon orbital spectrum for
 $\Lambda_{\rm QCD}$ = 0.22 GeV. Predictions for the nucleons are
 shown in (a) and for the $\Delta$ trajectories in (b).
 The lower dashed curves correspond to baryon states dual to
 spin-$\half$ modes in the bulk and the upper continuous curve to
 states dual to spin-$\threehalf$ modes.}
\label{fig:BaryonSpec}
\end{figure}

The field $\hat\Psi$ can be expanded in terms of eigenfunctions $\eta_\kappa(y)$
of the Dirac  operator on the compact
space $X$, $i \Dslash_X \eta_k(y) = \lambda_\kappa \eta_\kappa(y)$,
with eigenvalues $\lambda_\kappa$ as
$\hat\Psi(x, z, y) = \sum_\kappa \Psi_\kappa(x, z) \eta_\kappa(y)$,
where the $y$ are coordinates of $X$. The AdS Dirac equation
is~\cite{Muck:1998iz} 
\begin{widetext}
\vspace{-15pt}
\begin{equation}
\bigg[ z^2 ~\partial_z^2  - d~z~\partial_z + z^2  \mathcal{M}^2 -
 ( \lambda_\kappa + \mu)^2 R^2  y
 + \frac{d}{2} \left(\frac{d}{2} + 1\right) + (\lambda_\kappa + \mu)
 R~\hat\Gamma \bigg] f(z) = 0 ,
\end{equation}
\end{widetext}
where $\Psi(x,z) = e^{-i P \cdot x} ~ f(z)$ and
$\hat\Gamma u_\pm = \pm u_\pm$.
For $AdS_5$, $\hat\Gamma$ is the four-dimensional
chirality operator  $\gamma_5$. The AdS mass $\mu$ is determined
asymptotically by the orbital excitations in the boundary: $\mu =
L/R$. The eigenvalues on $S^{d+1}$ are
$\lambda_\kappa R = \pm \left(\kappa + \frac{d}{2} + \half \right)$,
$\kappa  = 0, 1, 2, ...$~\cite{Camporesi:1995fb}.
The normalizable modes for $\kappa = 0$ are
\begin{eqnarray}
\Psi_{\alpha,k}(x,z) =
C_{\alpha,k} e^{-i P \cdot x} z^{5/2}
[ J_\alpha(z \beta_{\alpha,k} \Lambda_{\rm QCD}  )~u_+(P) \\ \nonumber
+ J_{\alpha + 1} (z \beta_{\alpha,k} \Lambda_{\rm QCD} )~u_-(P) ] ,
\end{eqnarray}
where $u^-  = \frac{\gamma^\mu P_\mu}{P}~u^+$,
$\alpha = 2 + L$ and $\Delta = \frac{9}{2} + L$.
The solution of the spin-$\threehalf$ Rarita-Schwinger equation in AdS
space is more involved, but considerable simplification occurs in the
$\Psi_z = 0$ gauge for polarization
along Minkowski coordinates, $\Psi_\mu$, where it becomes
similar to the spin-$\half$ solution~\cite{Volovich:1998tj}.
The four-dimensional spectrum follows from
$\Psi^\pm(x,z_o) = 0$ or $\Psi^\pm_\mu(z,z_o) = 0$
\begin{equation} \label{eq:M+-}
\mathcal{M}_{\alpha, k}^+ = \beta_{\alpha,k} \Lambda_{\rm QCD}, ~~
\mathcal{M}_{\alpha, k}^- = \beta_{\alpha + 1,k} \Lambda_{\rm QCD},
\end{equation}
with a scale independent mass ratio.
Two of the fermions can be assigned to the fundamental
representation of $SU(N_C)$;  however,  to have a color-singlet state of
three fields at large $N_C$, the third must be in the $ N_C
(N_C-1)/2 $ antisymmetric representation~\cite{Dimopoulos:1980hn}.
For $N_C = 3$ we recover the usual interpolating operator
which creates a physical baryon in $QCD(3+1)$:
$\mathcal{O}_{9/2} = \epsilon_{a b c} \psi_a \psi_b \psi_c$.
The internal spin $S$ of a
given hadron matches the spin of its dual string.  The boundary conditions are $\Psi^+(x,
z_o) = 0$ for baryons with internal spin $S = \half$ and $\Psi^-_\mu(x, z_o) = 0$ for $S
=\threehalf$.  Figure \ref{fig:BaryonSpec}(a) shows the predicted orbital spectrum of the
nucleon states and Fig. \ref{fig:BaryonSpec}(b) the $\Delta$ orbital resonances.  The
only parameter is the value of $\Lambda_{\rm QCD}$ which we take as $0.22$ GeV. The baryon
states with internal spin $S = \half$ lie on a curve below the states with $S =
\threehalf$. The spectrum shows a clustering of states with the same orbital $L$,
consistent with a strongly suppressed spin-orbit force. Nucleon and $\Delta$ resonances
with same total quark spin $S$ fall on the same trajectory. With the exception of the
lowest states, the agreement of
the predicted spectrum with data is remarkable. However,
the first AdS radial state has a mass 1.85 GeV, so it is difficult to identify it
with the Roper $N\half^+(1440)$.

Eq. ({\ref{eq:M+-}) predicts a
novel parity degeneracy between states in the parallel trajectories shown in
Fig. \ref{fig:BaryonSpec}, as seen by displacing the
upper curve by one unit of $L$ to the right. Thus, the $L = 1$ states
$N(1650)$, $N(1675)$ and $N(1700)$ are degenerate with the $L = 2$,
$N(1680)$ and $N(1720)$, likewise the $L = 3$ states $N(2190)$ and
$N(2250)$ with the $L = 4$, $N(2220)$. The deltas provide another
excellent example of parity degeneracy: the $L = 2$ states $\Delta(1905)$,
$\Delta(1910)$, $\Delta(1920)$, $\Delta(1950)$ are within the error
bars degenerate with the $L = 3$ state $\Delta(1930)$.
It has been suggested that parity doublets with the same total angular
momentum, such as the $\Delta{\fivehalf^+}(1905)$ -
$\Delta{\fivehalf^-}(1930)$ doublet,
are due to chiral symmetry
restoration in the upper part of the light
baryon spectrum~\cite{Glozman:1999tk}, which is consistent
with the larger symmetry implied by  ({\ref{eq:M+-}).
In the quark-diquark model of Jaffe and Wilczek~\cite{Wilczek:2004im}, baryon
states on the lower trajectory of Fig. \ref{fig:BaryonSpec}(a),
correspond to ``good'' diquarks,  the upper to ``bad''
diquarks, and all the states shown in Fig. \ref{fig:BaryonSpec}(b) to ``bad''
diquarks, with exception of the $\Delta(1930)$
which does not follow the simple 3q quark-diquark pattern.

The general agreement of the holographic model with
the known light baryon spectrum is quite remarkable
and nontrivial.
The only mass scale in the holographic model is
$\Lambda_{\rm QCD}.$  The best fit to the meson spectra is $\Lambda_{\rm QCD}$
= 263 MeV; the best fit to the baryon spectrum is  $\Lambda_{\rm QCD}$ = 220
MeV. The small difference could be due to the different sensitivity of 
the mesons and baryons to the space truncations at $z_0 = 
1/\Lambda_{\rm QCD}$. Moreover,
the ratio of the delta to nucleon
trajectories is parameter independent, depending simply on the ratios
of zeros of Bessel functions.
Hadrons are identified  by
requiring that the state in the bulk has the correct matching
conformal dimension at $z \to 0, ~ x^2 \to 0.$
Our baryon analysis is thus based on color-singlet states which 
extrapolate to three fermion fields at zero separation.  
The contributions of higher particle Fock states of a hadron wave function
are suppressed by extra powers of $z$ at $z \to 0$,
so only the valence state is important in the short-distance domain.

The holographic model is relevant to
the color-singlet hadronic spectrum of any gauge theory which has
conformal scaling at short distances and confinement at large distances.
The degeneracy of the hadronic states depends on the flavor symmetry that is
assumed; {\it i.e.},  the number of massless quarks. There is no explicit dependence on $N_C$,
and  the QCD spectrum follows by matching dimensions to
$SU(3)_C$ color-singlet hadronic states at the $z\to 0$ boundary. 
For example, the 10-dimensional lowest Dirac AdS modes have dimension 9/2, 
precisely the conformal dimension of a $SU(3)_C$ 3-quark baryon state.
 
The SYM particles are expected to acquire a mass of the order of the supersymmetric
(SUSY) breaking scale and decouple from the theory. Consequently, the
only (non-supersymmetric) hadronic states which can be derived from the classical 
holographic theory are effectively the (dimension-$3$) $J^P=0^-,1^-$ pseudoscalar
and vector mesons, 
the (dimension-$\frac{9}{2}$) $J^P=\frac{1}{2}^+, \frac{3}{2}^+$ 
baryons, and the (dimension-$4$) $J^P= 0^+$ glueball states --  
corresponding exactly to the lowest-mass physical hadronic states. 
Our prediction for the mass of the lowest glueball state 
$\Theta^{++}$ is $\mathcal{M} \simeq  1.3$ GeV for $\Lambda_{\rm QCD} = 0.26$ GeV.
Hadrons with nonzero orbital angular momentum and higher Fock states require
the introduction of quantum fluctuations. The model provides an explanation
of why hadrons consist of two gluons, three quarks, or a quark and antiquark,
and not other exotic combinations. 

In some sense the holographic model is a
covariant generalization of the MIT/SLAC bag models;  however, unlike
the bag models, it also incorporates the 
near-conformal behavior of QCD at short distances.
The approach is highly successful in organizing the hadron spectrum, 
although it  underestimates the spin-orbit
separations of the $L = 1$ orbital states.
The model might be improved for the low-lying states by modifying 
the boundary conditions at $r=r_0$.
Our results suggest that basic
features of the QCD hadron spectrum can be understood in terms
of a higher dimensional dual theory.

\begin{acknowledgments}

We thank James D. Bjorken, Elena Caceres, Simon Capstick, Lance Dixon,
Leonid Glozman, Joe Polchinski,
David Richards and Matt Strassler for helpful comments.
This research was supported in part by the National Science Foundation
under Grant No. PHY99-07949 and by the Department
of Energy contract DE--AC02--76SF00515.

\end{acknowledgments}

\end{document}